\documentclass[10pt,aps,showpacs,a4paper,floatfix,onecolumn]{revtex4}
\usepackage{vmargin}
\usepackage[T1]{fontenc}
\usepackage{epsfig}
\usepackage{psfig}
\usepackage{graphicx}
\usepackage{graphics}
\usepackage{xspace}
\usepackage{amssymb}
\usepackage{amsmath}
\usepackage{latexsym}
\usepackage{natbib}
\usepackage{mathrsfs}
\newcommand{\diff}{{\rm\,d}}                    
\newcommand{\dida}{\footnotesize}               
\newcommand{\ove}{\overline}                    
\def\fps@figure{htb}
\def\fps@table{htb}

\def\r{\mbox{\boldmath $r$}}

\def\p{\mbox{\boldmath $p$}}

\def\q{\mbox{\boldmath $q$}}

\def\ss{\mbox{\boldmath $\sigma$}}

\def\r{\mbox{\boldmath $r$}}
\def\pf{\mbox{\boldmath $p^{ \prime}$}}
\def\pm{\mbox{\boldmath $p_{\mathrm m}$}}

\def\ep{\mbox{$e^{\prime}$}}
\mathchardef\varepsilon="010F
\mathchardef\epsilon="0122
\mathchardef\theta="0123
\mathchardef\vartheta="0112
\begin{document}

\bibliographystyle{apsrev}

\title{Relativistic calculation of nuclear transparency in 
$\left(e,\ep p\right)$ reactions}
\author{Andrea Meucci}
\affiliation{Dipartimento di Fisica Nucleare e Teorica, Universit\`a di 
Pavia} 
\affiliation{ Istituto Nazionale di Fisica Nucleare, Sezione di Pavia, Italy}
\date{\today}

\begin{abstract}
\vskip .5cm
Nuclear transparency in $\left(e,\ep p\right)$ reactions is evaluated in a fully
relativistic distorted wave impulse approximation model. The relativistic mean 
field theory is used for the bound state and the Pauli reduction for the 
scattering state, which is calculated from a relativistic optical potential. 
Results for selected nuclei are displayed in a $Q^2$ range between $0.3$ and 
$1.8$ (GeV$/c)^2$ and compared with
recent electron scattering data. For $Q^2 = 0.3$ (GeV$/c)^2$ the results are
lower than data; for higher $Q^2$ they are in reasonable agreement with
data. The sensitivity of the model to
different prescriptions for the one-body current operator is investigated. The 
off-shell ambiguities are rather large for the distorted cross sections and 
small for the plane wave cross sections.
\end{abstract}


\maketitle


\section{Introduction}
\label{intro}

Exclusive $\left(e,\ep p\right)$ knockout reactions have been used since a long
time to study the single particle properties of nuclear structure. The analysis
of the experimental cross sections were successfully carried out in the
theoretical framework of the nonrelativistic distorted wave impulse
approximation (DWIA) for $Q^2$ less than $0.4$ GeV$/c^2$~\cite{Oxford,Kelly1}. 
In recent years, owing to the new data from TJNAF~\cite{gao,malov}, similar 
models
based on a fully relativistic DWIA (RDWIA) framework were developed. In this
approach the wave functions of the initial and final nucleons are solutions of a
Dirac equations containing scalar and vector potentials fitted to the ground
state properties of the nucleus and to proton elastic scattering
data~\cite{RDWIA}.

In the nucleus, final state interaction with the nuclear medium can absorb the
struck proton, thus reducing the experimental cross section. This reduction is
related to the nuclear transparency, which can be intuitively defined as the
ratio of the measured to the plane wave cross section. The transparency can be
used to refine our knowledge of nuclear medium effects and to look for deviation
from conventional predictions of nuclear physics, such as the Color Transparency
(CT) effect. The CT was introduced basing on  perturbative QCD
arguments~\cite{ct}. The name is related to the disappearance of the color
forces at high $Q^2$: three quarks should form an object that passes through the
nuclear medium without undergoing interactions. If the CT effect switches on as 
$Q^2$ increases, then the nuclear transparency should be enhanced towards unity. 
Several measurements of the nuclear transparency in $\left(p,2p\right)$ and
$\left(e,\ep p\right)$ knockout have been carried out in the past. The first
experiment looking for CT effect was performed at Brookhaven~\cite{brook}
measuring transparency in $\left(p,2p\right)$ reaction. An increase of
transparency for $3\leq Q^2\leq 8$ (GeV/$c)^2$, followed by a decrease for
$8\leq Q^2\leq 11$ (GeV/$c)^2$ was observed. New data confirm this energy
dependence of transparency~\cite{brook2}. The first measurements of nuclear
transparency in $\left(e,\ep p\right)$ reaction were carried out at Bates with
$Q^2 = 0.3$ (GeV$/c)^2$~\cite{garino}. In recent years, higher energy data
of transparency in $\left(e,\ep p\right)$ were produced at SLAC~\cite{oneill}
and TJNAF~\cite{abbott,garrow}. In contrast with $\left(p,2p\right)$ data, the
NE-18 experiment at SLAC did not see any CT effect up to $Q^2 =6.8$ (GeV$/c)^2$,
but could not exclude a slow onset of CT. The E91-013 experiment at TJNAF
studied the nuclear transparency in a $Q^2$ range up to $8.1$ (GeV$/c)^2$ with
greatly improved statistics and did not found evidence for the onset of CT.

The distorted wave approach was first applied to evaluate transparency in
$\left(e,\ep p\right)$ knockout in Ref.~\cite{green}, where it was shown
that measurements of the normal transverse structure function in $^{208}$Pb 
could afford to see CT effect, and in Ref.~\cite{kellyt}, where the nuclear 
part of the transition amplitude was written in terms of Schr\"odinger-like
wave functions for bound and scattering states and of an effective current
operator containing the Dirac potentials. Alternatively, the nuclear
transparency results were analyzed in terms of a Glauber
model~\cite{pand,jain,frank1}, which assumes classical attenuation of protons 
in the nuclear medium.

In this paper we present RDWIA calculations of nuclear transparency in 
$\left(e,\ep p\right)$ reaction.
The RDWIA treatment is the same as in Refs.~\cite{meucci1,meucci2}. The 
relativistic 
bound state wave functions have been generated as solutions of a Dirac equation 
containing scalar and vector potentials obtained in the framework of the 
relativistic mean field theory.
The effective Pauli reduction has been adopted for the outgoing nucleon
wave function. The resulting Schr\"odinger-like equation is
solved for each partial wave starting from relativistic optical potentials.
The relativistic current is written 
following the most commonly used current conserving $(cc)$ prescriptions 
for the ($e,e'p$) reaction introduced in Ref.~\cite{deF}. The ambiguities 
connected with different choices of the electromagnetic current cannot generally
be dismissed. In the $\left(e,\ep p\right)$ reaction the predictions of 
different 
prescriptions are generally in close agreement~\cite{pollock}. Large differences
can however be found at high missing momenta~\cite{off1,off2}. 

The formalism is outlined in Sec.~\ref{formalism}. Relativistic 
calculations of nuclear transparency are presented in Sec.~\ref{results}, where 
current ambiguities are also investigated. Some conclusions are drawn in 
Sec.~\ref{con}.

\section{Formalism}
\label{formalism}

The nuclear transparency can be experimentally defined as the ratio of the
measured cross section to the cross section in plane wave approximation, which
is usually evaluated by means of a Monte Carlo simulation to take in account the
kinematics of the experiment.
Hence, we define nuclear transparency as
\begin{eqnarray}
T = \frac {\int_V \diff E_{\mathrm m} \diff\pm~\sigma_{DW} \left( 
E_{\mathrm m}, \pm, \pf\right)}{\int_V \diff E_{\mathrm m}\diff\pm~ 
\sigma_{PW} \left(E_{\mathrm m}, \pm\right)} \ , \label{eq.transpa}
\end{eqnarray} 
where $\sigma_{DW}$ is the distrorted wave cross section and $\sigma_{PW}$ is 
the plane wave one.
Since the measured transparency depends upon the kinematics
conditions and the spectrometer acceptance, we have to specify the space 
phase volume, $V$, and use it for both the numerator and the
denominator~\cite{gol}.
Because of final state interaction, the distorted cross section depends upon 
the momentum of the emitted nucleon $\pf$, whereas the undistorted cross section
only depends upon the missing energy $E_{\mathrm m}$ and the missing momentum
$\pm$.  

In the one-photon exchange approximation the $\left(e,\ep p\right)$ cross 
section is given by the contraction between the
lepton tensor and the hadron tensor. In the case of an unpolarized reaction it
can be written as
\begin{eqnarray}
\sigma  = \sigma _{\mathrm M}\ f_{\mathrm {rec}}\ E' |\pf |\ \left[\rho _{00}
f_{00}+  \rho _{11}f_{11}+\rho _{01}f_{01}\cos\left(\alpha\right)+
  \rho _{1-1}f_{1-1}\cos\left(2\alpha\right)\right] \ ,  \label{eq.fcs}
\end{eqnarray}
where $\sigma _{\mathrm M}$ is the Mott cross section, $f_{\mathrm {rec}}$ is 
the recoil factor \cite{Oxford,Kelly1}, $E'$ and $\pf$ are the energy and 
momentum of the emitted nucleon, and $\alpha $ is the out of plane angle
between the electron scattering plane and the $(\q, \pf)$ plane. The 
coefficients $\rho_{\lambda\lambda'}$ are obtained from the lepton tensor
components and depend only upon the electron kinematics \cite{Oxford,Kelly1}. 
The structure functions $f_{\lambda\lambda'}$ are given by bilinear combinations
of the components of the nuclear current as
\begin{eqnarray}
f_{00} &=& \langle J^0 \left(J^0\right)^{\dagger }\rangle \ , \nonumber \\ 
f_{11} &=& \langle J^x \left(J^x\right)^{\dagger} \rangle +
          \langle J^y \left(J^y\right)^{\dagger} \rangle \ , \nonumber \\
f_{01} &=& -2\sqrt 2\ \mathfrak{Re}
   \left[\langle J^x \left(J^0\right)^{\dagger }\rangle\right] \ , \nonumber \\
f_{1-1} &=&  \langle J^y \left(J^y\right)^{\dagger} \rangle -
          \langle J^x \left(J^x\right)^{\dagger} \rangle \ ,
\end{eqnarray}
where $\langle\cdots\rangle$ means that average over the initial and sum over
the final states is performed fulfilling energy conservation. 
In our frame of reference the $z$ axis is along $\q$, and the $y$ axis is
parallel to $\q\times\pf$.

In RDWIA the matrix elements of the nuclear current operator, i.e.,  
\begin{eqnarray}
J^{\mu} = \int \diff \r \ove \Psi_f(\r) \widehat j^{\mu} \exp{\{i\q\cdot \r\}}
 \Psi_i(\r) \ , \label{eq.rj}
 \end{eqnarray}
are calculated using relativistic wave functions for initial and final states.

The choice of the electromagnetic operator is a longstanding problem. Here
we discuss the three current conserving expressions
\cite{deF,Kelly2,Kelly3}
\begin{eqnarray}
\widehat j_{cc1}^{\mu} &=& G_M(Q^2) \gamma ^{\mu} - 
             \frac {\kappa}{2M} F_2(Q^2)\overline P^{\mu} \ , \nonumber \\
\widehat j_{cc2}^{\mu} &=& F_1(Q^2) \gamma ^{\mu} + 
             i\frac {\kappa}{2M} F_2(Q^2)\sigma^{\mu\nu}q_{\nu} \ ,
	     \label{eq.cc} \\
\widehat j_{cc3}^{\mu} &=& F_1(Q^2) \frac{\overline P^{\mu}}{2M} + 
             \frac {i}{2M} G_M(Q^2)\sigma^{\mu\nu}q_{\nu} \ , \nonumber
\end{eqnarray}
where $q^{\mu} = (\omega,\q)$ is the four-momentum transfer,
$Q^2=\mid\q\mid^2-\omega ^2$, $\overline P^{\mu} = (E+E',\pm+\pf)$, 
$\kappa$ is the anomalous part of the magnetic
moment, $F_1$ and $F_2$ are the Dirac and Pauli nucleon form factors, $G_M =
F_1+\kappa F_2$ is the Sachs nucleon magnetic form factor, and
$\sigma^{\mu\nu}=\left(i/2\right)\left[\gamma^{\mu},\gamma^{\nu}\right]$. These 
expressions are equivalent for on-shell particles thanks to Gordon identity. 
However, since nucleons in the nucleus are off-shell we expect that these 
formulas should give different results. Current 
conservation is restored by replacing the longitudinal current and the bound 
nucleon energy by  \cite{deF}
\begin{eqnarray}
J^L &=& J^z = \frac{\omega}{\mid\q\mid}~J^0 \ , \\
E &=& \sqrt{\mid \pm \mid^2 + M^2} = \sqrt{ \mid \pf-\q\mid^2 + M^2} \ .
\end{eqnarray}

The bound state wave function
\begin{eqnarray}
\Psi_i = \left(\begin{array}{c} u_i \\ v_i \end{array}\right) \ , \label{eq.bwf}
\end{eqnarray}
is given by the Dirac-Hartree solution of a relativistic Lagrangian
containing scalar and vector potentials. 

The ejectile wave function $\Psi_f$ is written in terms of its positive energy
component $\Psi_{f+}$ following the direct Pauli reduction method~\cite{HPa}
\begin{eqnarray}
\Psi_f = \left(\begin{array}{c} \Psi_{f+} \\ \frac {\ss\cdot\p'}{M+E'+S-V}
        \Psi_{f+} \end{array}\right) \ ,
\end{eqnarray}
where $S=S(r)$ and $V=V(r)$ are the scalar and vector potentials for the nucleon
with energy $E'$. The upper component $\Psi_{f+}$ is related to a
Schr\"odinger equivalent wave function $\Phi_{f}$ by the Darwin factor $D(r)$,
i.e.,
\begin{eqnarray}
\Psi_{f+} &=& \sqrt{D(r)}\Phi_{f} \ , \\
D(r) &=& \frac{M+E'+S-V}{M+E'} \ .
\end{eqnarray}
$\Phi_{f}$ is a two-component wave function which is solution of a 
Schr\"odinger
equation containing equivalent central and spin-orbit potentials obtained from
the scalar and vector potentials. Hence, using the relativistic 
normalization, the emitted nucleon wave function is written as
\begin{eqnarray}
\ove \Psi _f = \Psi _f^{\dagger}\gamma ^0 &=& \sqrt {\frac {M+E'}{2E'}}\
\left[ \left(\begin{array}{c} 1 \\ \frac {\ss \cdot \p'}{C}\end{array} \right) 
\sqrt {D}\ \Phi _f \right] ^{\dagger }\ \gamma ^0  \nonumber \\
&=& \sqrt {\frac {M+E'}{2E'}}\ \Phi _f^{\dagger }
\left(\sqrt {D}\right) ^{\dagger } \left( 1\ ;\ \ss \cdot \p' 
\frac {1}{C^{\dagger }}\right) \gamma ^0  \ , \label{eq.psif}
\end{eqnarray}
where 
\begin{eqnarray}
C = C(r) = M + E' + S(r) - V(r) \ . \label{eq.cf}
\end{eqnarray}

\section{Transparency and the \lowercase{$\left(e,\ep p\right)$} reaction}
\label{results}

The $\left(e,\ep p\right)$ reaction is a well-suited process to search for CT
effects. The $e$-$p$ cross section is accurately known from QED and the energy
resolution guarantees the exclusivity of the reaction.
Several measurements of nuclear transparency to protons in quasifree 
$\left(e,\ep p\right)$ knockout have been carried out  
on several target nuclei and over a wide range of energies to look for CT onset. 

Here, we calculated nuclear transparency for closed shell or subshell nuclei 
at kinematics conditions compatible with the experimental setups for which the
measurements of nuclear transparency have been performed, and for which the
RDWIA predictions are known to provide a good agreement with cross section data. 
The bound state wave functions and optical potentials are the same as in 
Refs.~\cite{meucci1,meucci2}, where the RDWIA results are in satisfactory 
agreement with $\left(e,\ep p\right)$ and $\left(\gamma,p\right)$ data.

The relativistic bound-state wave functions have 
been obtained from the program ADFX of Ref.~\cite{adfx}, where relativistic 
Hartree-Bogoliubov equations are solved in the mean field approximation to the 
description of ground state properties of several spherical nuclei. 
The model starts from a Lagrangian density containing
sigma-meson, omega-meson, rho-meson and photon field, whose potentials are 
obtained by solving self-consistently Klein-Gordon 
equations. Moreover, finite range interactions are included to describe pairing
correlations and the coupling to particle continuum states.

The outgoing nucleon wave function is calculated by means of the complex
phenomenological optical potential EDAD1 of Ref.~\cite{chc}, which is obtained 
from fits to proton elastic scattering data on several nuclei in an energy 
range up to 1040 MeV.  

Since no rigorous prescription exists for handling off-shell nucleons,
we have studied the sensitivity to different $cc$ choices of the 
nuclear current. The Dirac and Pauli form factors $F_1$ and $F_2$ are taken 
from Ref.~\cite{mud}. 

\begin{figure}[h]
\begin{center}
\includegraphics[bb=40 140 600 750,scale=.85]{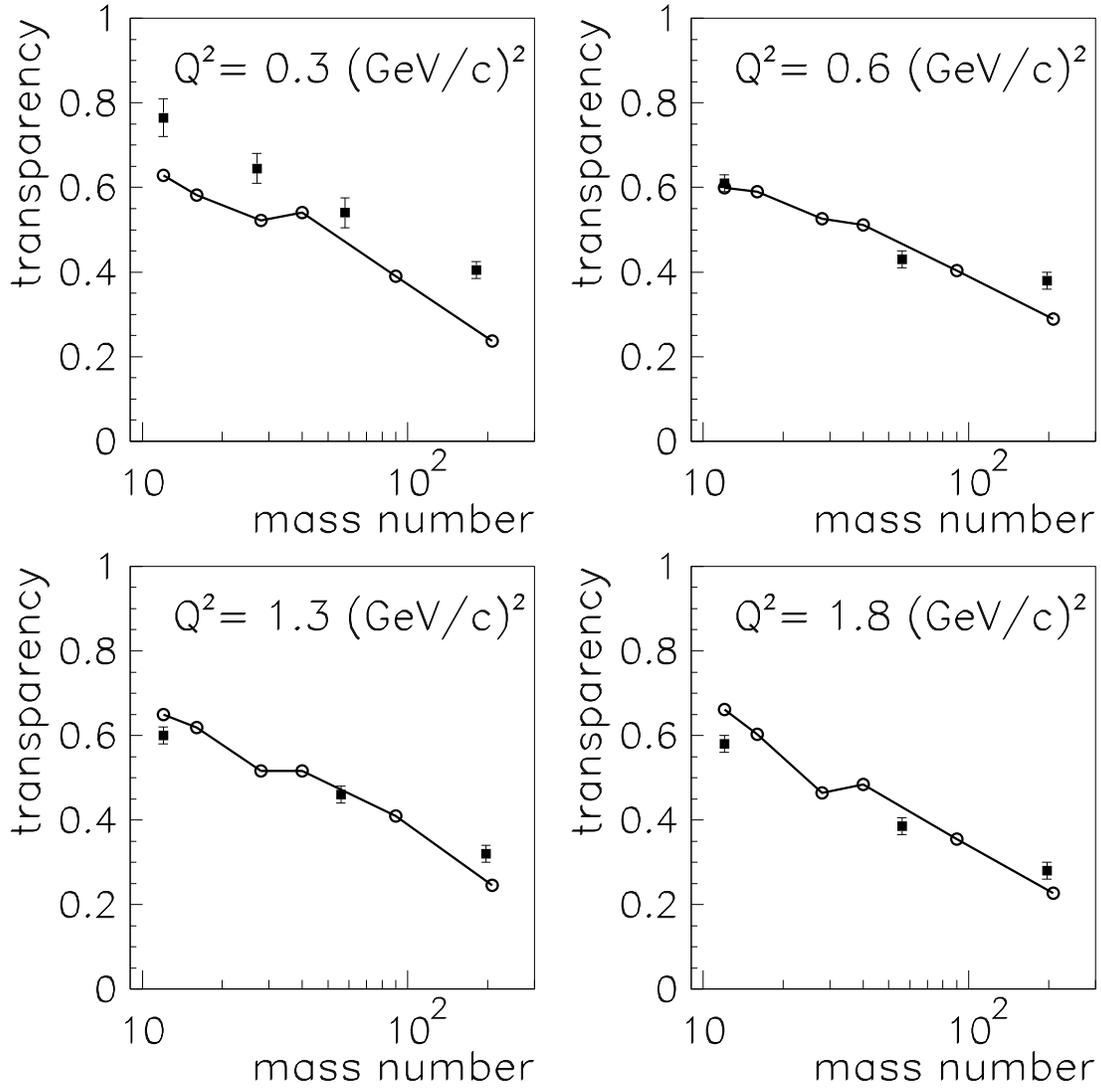}
\caption {\dida {The nuclear transparency for the quasifree 
A$\left(e,\ep p\right)$ reaction as a function of the mass number for $Q^2$
ranging from $0.3$ to $1.8$ (GeV$/c)^2$. Calculations have been performed for
selected closed shell or subshell nuclei with mass numbers indicated by open 
circles. The data at $Q^2 = 0.3$ (GeV$/c)^2$ are from Ref.~\cite{garino}. The 
data at $Q^2 = 0.6, 1.3$, and $1.8$ (GeV/$c)^2$ are from Ref.~\cite{abbott}.}
\label{fig.ta}}
\end{center}

\end{figure}

In Fig.~\ref{fig.ta} our RDWIA results for nuclear transparency, calculated with
the $cc2$ prescription for the nuclear current are shown. The $Q^2$ of the 
exchanged photon is taken between $0.3$ (GeV/$c)^2$ and $1.8$ (GeV/$c)^2$ in 
constant $\left(\q, \omega\right)$ kinematics. Calculations have been performed 
for selected closed shell or subshell nuclei ($^{12}$C, $^{16}$O, $^{28}$Si, 
$^{40}$Ca, $^{90}$Zr, and $^{208}$Pb) for which the relativistic mean field code
easily converges. The agreement with the data is rather satisfactory. At 
$Q^2 = 0.3$ (GeV/$c)^2$ our results lie below the data and are comparable with
those presented in Ref.~\cite{kellyt}, where it was shown that the EDAD1 optical
potential led to a smaller transparency, while better agreement was found using
an empirical effective interaction which fits both proton elastic and inelastic 
scattering data. However, we have to note that the DWIA model of 
Ref.~\cite{kellyt} uses a different approach to obtain single particle bound
state wave functions. The calculations at $Q^2 = 0.6, 1.3$, and $1.8$ 
(GeV/$c)^2$ are closer to the data and fall down only for higher mass numbers.

\begin{figure}[h]
\begin{center}
\includegraphics[bb=40 140 600 750,scale=.85]{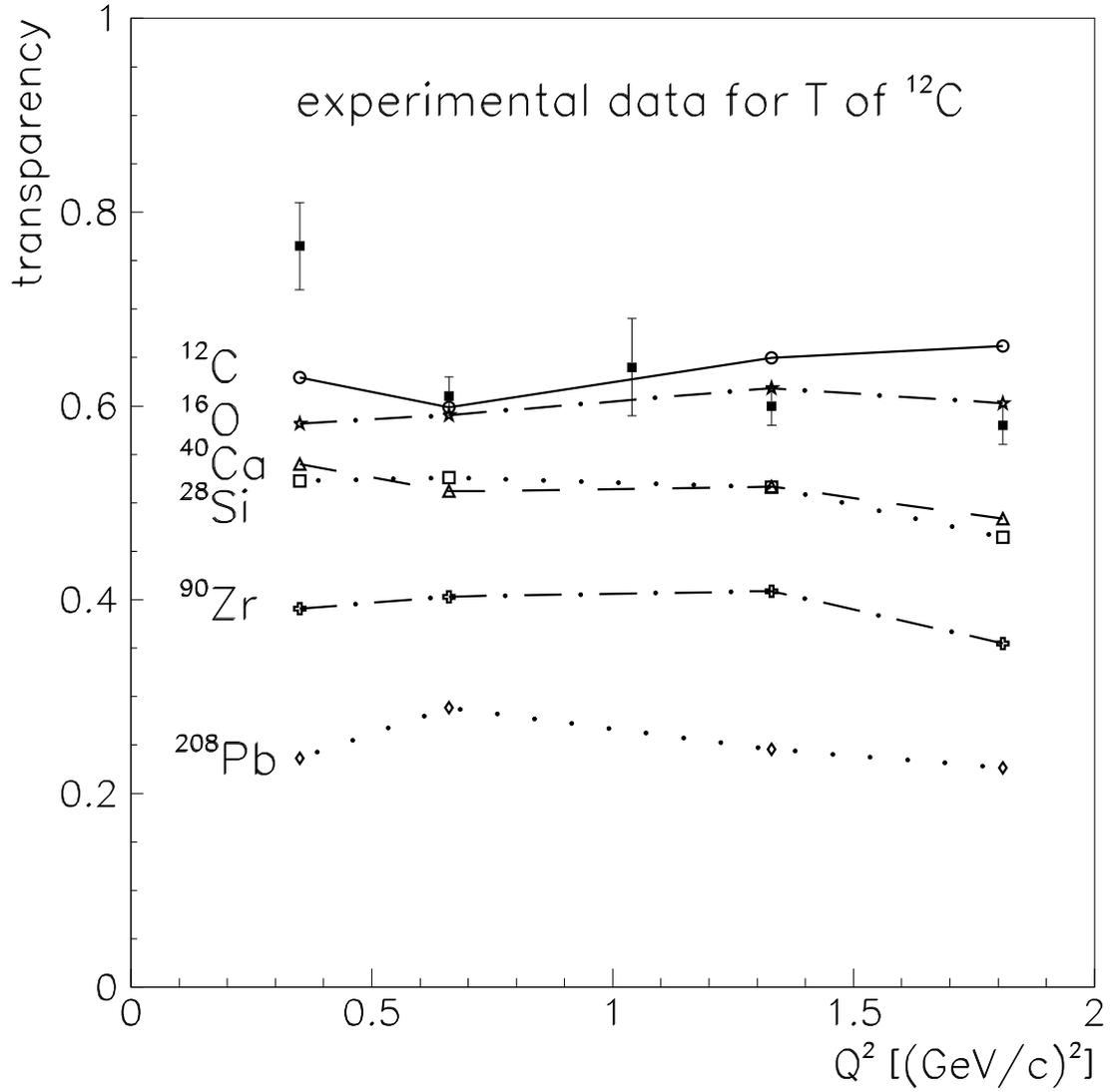}
\caption {\dida {The energy dependence of nuclear transparency for $^{12}$C
(open circles), $^{16}$O (open stars), $^{28}$Si (open squares), $^{40}$Ca 
(open triangles), $^{90}$Zr (open crosses), and $^{208}$Pb(open diamonds), 
at the same kinematics as in Fig.~\ref{fig.ta}. Calculations were performed for
$Q^2$ values marked by symbols. The $^{12}$C data are from 
Refs.~\cite{garino,oneill,abbott}.}\label{fig.tnuc}}
\end{center}
\end{figure}

In Fig.~\ref{fig.tnuc} the energy dependence of nuclear transparency is shown.
The calculations have been performed for the same nuclei and at the same
kinematics as in Fig.~\ref{fig.ta}. The calculated transparency is approximately
constant for each nucleus and decreases for increasing mass number. 

In Refs.~\cite{oneill,garrow} it is reported that the transparency data can be
fitted with an exponential law of the form $T=A^{-\alpha}$, with $\alpha\simeq
0.24$. Since our model is based on a single particle 
picture of nuclear structure, we expect our results to be sensible to the
discontinuities of the shell structure. These clearly appear in the changes in
shape of the A-dependent curves.

\begin{figure}[h]
\begin{center}
\includegraphics[bb=40 140 600 750,scale=.85]{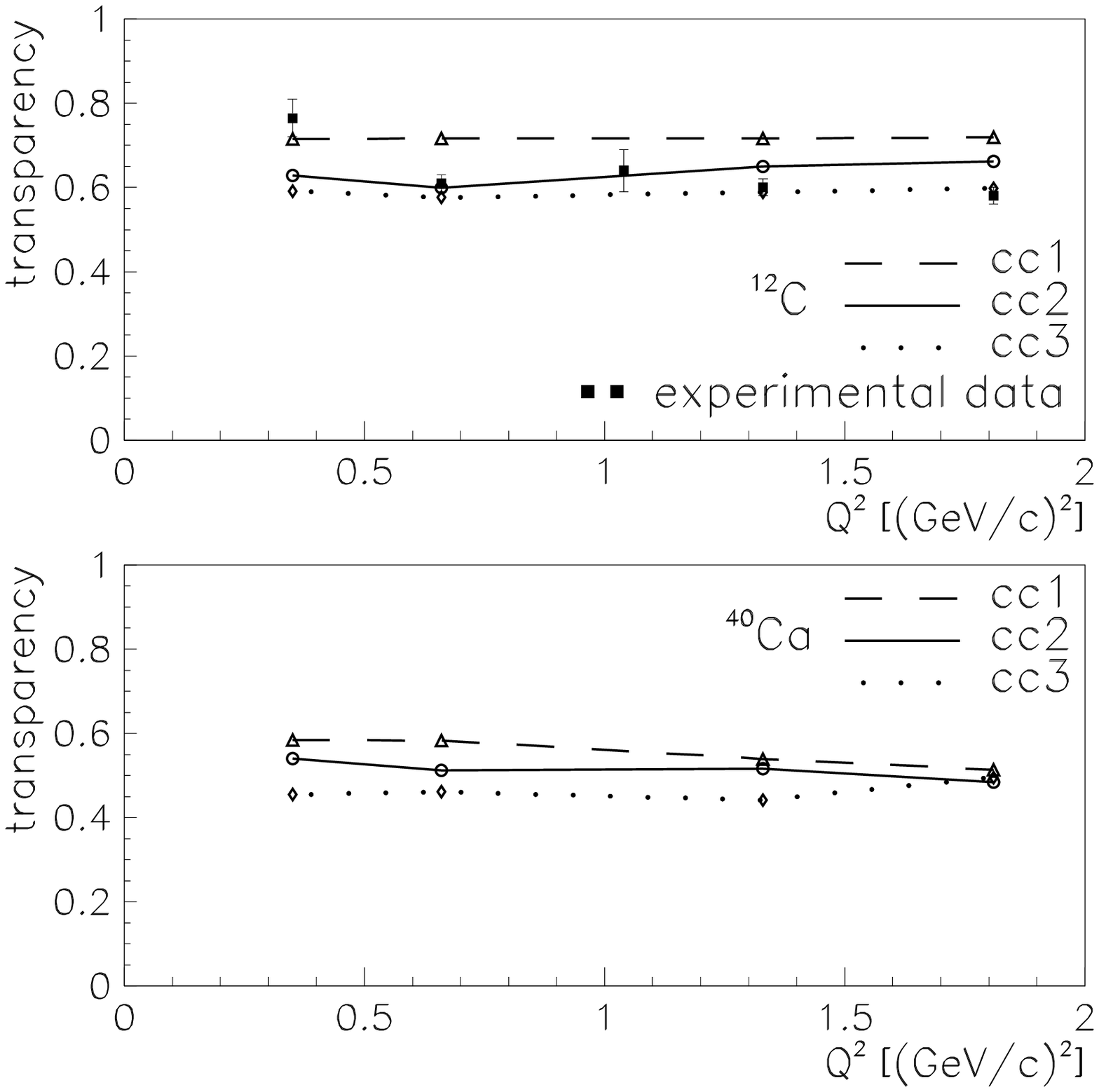}
\caption {\dida {The electromagnetic current dependence of nuclear transparency 
for $^{12}$C and $^{40}$Ca, at the same kinematics as in
Fig.~\ref{fig.ta}. Calculations were performed for
$Q^2$ values marked by symbols.}\label{fig.tcc}}
\end{center}
\end{figure}

In Fig.~\ref{fig.tcc} the sensitivity of transparency calculations for $^{12}$C
and $^{40}$Ca to different choices for the electromagnetic current is shown. 
The results with the $cc1$ current are larger than those
obtained with the $cc2$ current, whereas $cc3$ results are smaller than the 
$cc2$ ones. A similar behavior was already found out in Ref.~\cite{meucci2} for
$\left(\gamma,N\right)$ differential cross section. Here it 
is mainly due to the fact that, when using the $cc1$ current, the 
distorted cross section, $\sigma_{DW}$ in Eq.~\ref{eq.transpa}, is enhanced 
with respect to the calculations with the $cc2$ or the $cc3$ current, whereas 
the plane wave cross sections, $\sigma_{PW}$, are almost independent of the 
operator form.

\section{Summary and conclusions}
\label{con}

In this paper we have presented relativistic DWIA 
calculations for nuclear transparency of $\left(e,\ep p\right)$ reaction in a
momentum transfer range between $0.3$ and $1.8$ (GeV$/c)^2$. 

The transition matrix element of the nuclear current operator in RDWIA is
calculated using the bound state wave functions obtained in the framework of the 
relativistic mean field theory, and the direct Pauli reduction method with 
scalar and vector potentials for the scattering state. In order to analyze 
the ambiguities in the choice of the electromagnetic vertex due to the off-shell 
character of the initial nucleon, we have used three current conserving 
expressions in our calculations. 

We have performed calculations for selected closed shell or subshell nuclei. 
The dependence of nuclear transparency upon the mass number and the energy 
has been discussed. Low $Q^2$ results underestimates the data, thus indicating 
the presence of too strong an absorptive term in the optical potential. In 
contrast, results at higher $Q^2$ are closer to the data. We find little 
evidence of energy dependence or momentum transfer of the transparency for each 
nucleus.

The sensitivity to different choices of the nuclear current has been
investigated for $^{12}$C and $^{40}$Ca. 
The results with the $cc1$ current are larger than the $cc2$ results, whereas 
those obtained with the $cc3$ current are more similar to the $cc2$ ones. This 
effect is due to the enhancement of the $cc1$ distorted cross section with 
respect to the $cc2$ and $cc3$ cross sections.

\begin{acknowledgments}
I would like to thank Professor C.~Giusti and Professor F.~D.~Pacati for useful 
discussions and comments on the manuscript.
\end{acknowledgments}

%

 
%
%
%
%
%
%


\begin{thebibliography}{100}

\bibitem{Oxford}
S.~Boffi, C.~Giusti, F.~D.~Pacati, and M.~Radici, 
\newblock {\it Electromagnetic Response of Atomic Nuclei}, Oxford Studies in 
Nuclear Physics (Clarendon Press, Oxford, 1996).

\bibitem{Kelly1}
J.~J.~Kelly, 
\newblock Adv. Nucl. Phys. {\bf 23}, 75 (1996).

\bibitem{gao}
J.~Gao {\it et al.},
\newblock Phys. Rev. Lett. {\bf 84}, 3265 (2000).

\bibitem{malov}
S.~Malov {\it et al.},
\newblock Phys. Rev. C {\bf 62}, 057302 (2000).

\bibitem{RDWIA}
Y.~Jin, D.~S.~Onley, and L.~E.~Wright, 
\newblock Phys. Rev. C {\bf 45}, 1311 (1992);\\
Y.~Jin and D.~S.~Onley, 
\newblock Phys. Rev. C {\bf 50}, 377 (1994);\\
J.~M.~Ud\'{\i}as, P.~Sarriguren, E.~Moya de Guerra, E.~Garrido, and 
J.~A.~Caballero, 
\newblock Phys. Rev. C {\bf 48}, 2731 (1993);\\
J.~M.~Ud\'{\i}as, P.~Sarriguren, E.~Moya de Guerra, E.~Garrido, and 
J.~A.~Caballero, 
\newblock Phys. Rev. C {\bf 51}, 3246 (1995);\\
J.~M.~Ud\'{\i}as, P.~Sarriguren, E.~Moya de Guerra, and J.~A.~Caballero, 
\newblock Phys. Rev. C {\bf 53},  R1488 (1996);\\
J.~M.~Ud\'{\i}as and J.~R.~Vignote, 
\newblock Phys. Rev. C {\bf 62}, 034302 (2000).

\bibitem{ct}
A.~H.~Mueller, 
\newblock in {\it Proceedings of the XVII Rencontre de Moriond, 1982},
edited by J.~Tran Thanh Van (Editions Frontieres, Gif-sur-Yvette, France, 1982),
p.~13;\\
S.~J.~Brodsky,
\newblock in {\it Proceedings of the Thirteenth International Symposium on
Multiparticle Dynamics}, edited by W.~Kittel, W.~Metzger, and A.~Stergiou (World
Scientific, Singapore, 1982), p.~963.

\bibitem{brook}
A.~S.~Carroll {\it et al.},
\newblock Phys. Rev. Lett. 61, 1698 (1988).

\bibitem{brook2}
A.~Leksanov {\it et al.},
\newblock Phys. Rev. Lett. 87, 212301 (2001).

\bibitem{garino}
G.~Garino {\it et al.},
\newblock Phys. Rev. C {\bf 45}, 780 (1992).

\bibitem{oneill}
T.~G.~O'Neill {\it et al.},
\newblock Phys. Lett. {\bf B351}, 87 (1995).

\bibitem{abbott}
D.~Abbott {\it et al.},
\newblock Phys. Rev. Lett. {\bf 80}, 5072 (1998).

\bibitem{garrow}
K.~Garrow {\it et al.},
arXiv:hep-ex/0109027.

\bibitem{green} 
W.~R.~Greenberg and G.~A.~Miller,
\newblock Phys. Rev. C {\bf 49}, 2747 (1994).

\bibitem{kellyt}
J.~J.~Kelly, 
\newblock Phys. Rev. C {\bf 54}, 2547 (1996). 

\bibitem{pand}
V.~R.~Pandharipande and S.~C.~Pieper,
\newblock Phys. Rev. C {\bf 45}, 791 (1992). 

\bibitem{jain}
P.~Jain and J.~P.~Ralston,
\newblock Phys. Rev. D {\bf 48}, 1104 (1993).

\bibitem{frank1}
L.~Frankfurt, M.~Strikman, and M.~Zhalov,
\newblock Phys. Lett. {\bf B503}, 87 (2001).

\bibitem{meucci1}
A.~Meucci, C.~Giusti, and F.~D.~Pacati, 
\newblock Phys. Rev. C {\bf 64}, 014604 (2001).

\bibitem{meucci2}
A.~Meucci, C.~Giusti, and F.~D.~Pacati, 
\newblock Phys. Rev. C {\bf 64}, 064615 (2001).

\bibitem{deF}
T.~de Forest, Jr., 
\newblock Nucl. Phys. {\bf A392}, 232 (1983). 

\bibitem{pollock} 
S.~Pollock, H.~W.~L.~Naus, and J.~H.~Koch,
\newblock Phys. Rev. C {\bf 53}, 2304 (1996).
     
\bibitem{off1}
J.~A.~Caballero, T.~W.~Donnelly, E.~Moya de Guerra, and J.~M.~Ud\'{\i}as,
\newblock Nucl. Phys.   {\bf A632}, 323 (1998).

\bibitem{off2}
S.~Jeschonnek and J.~W.~Van Orden, 
\newblock Phys. Rev. C {\bf 62}, 044613 (2000).

\bibitem{gol} 
Y.~S.~Golubeva, L.~A.~Kondratyuk, A.~Bianconi, S.~Boffi, and M.~Radici,
\newblock Phys. Rev. C {\bf 57}, 2618 (1998).

\bibitem{Kelly2}
J.~J.~Kelly, 
\newblock Phys. Rev. C {\bf 56}, 2672 (1997); {\bf 59}, 3256 (1999). 

\bibitem{Kelly3}
J.~J.~Kelly, 
\newblock Phys. Rev. C {\bf 60}, 044609 (1999).

\bibitem{HPa}
M.~Hedayati-Poor, J.~I.~Johansson, and H.~S.~Sherif,
\newblock Nucl. Phys. {\bf A593}, 377 (1995); \\
M.~Hedayati-Poor, J.~I.~Johansson, and H.~S.~Sherif,
\newblock Phys. Rev. C {\bf 51}, 2044 (1995).

\bibitem{adfx} 
W.~P\"oschl, D.~Vretenar, and P.~Ring,
\newblock Comput. Phys. Commun. {\bf 103}, 217 (1997).

\bibitem{chc} E.~D.~Cooper, S.~Hama, B.~C.~Clark, and R.~L.~Mercer,
\newblock Phys. Rev. C {\bf 47}, 297 (1993).

\bibitem{mud} 
P.~Mergell, Ulf-G.~Meissner, and D.~Drechsel,
\newblock Nucl. Phys. {\bf A596}, 367 (1996).

\end{thebibliography}
\end{document}